\documentclass[aps,pre,reprint,groupedaddress,amssymb]{revtex4-1}

\makeatletter
\def\set@curr@file#1{%
  \begingroup
    \escapechar\m@ne
    \xdef\@curr@file{\expandafter\string\csname #1\endcsname}%
  \endgroup
}
\def\quote@name#1{"\quote@@name#1\@gobble""}
\def\quote@@name#1"{#1\quote@@name}
\def\unquote@name#1{\quote@@name#1\@gobble"}
\makeatother
\usepackage{graphicx}
\usepackage{dcolumn}
\usepackage{bm}
\usepackage{amsmath}
\usepackage{xcolor}
\usepackage{hyperref}
\usepackage{nicefrac}
\usepackage{caption}
\usepackage{subcaption}
\usepackage{epstopdf}
\usepackage{tikz}
\usepackage{caption}
\usepackage{nicefrac}
\usepackage{subfloat}
\usepackage{float}
\usepackage[margin=10pt,font=small,labelfont=bf,justification=centerlast,format=plain]{caption}

\begin{document}

\newcommand{\RE}{\mathrm{Re}}
\newcommand{\IM}{\mathrm{Im}}
\title{Mean-Field Solution for Critical Behavior of Signed Networks in Competitive Balance Theory}

\author{R. Masoumi}
\author{F. Oloomi}
\author{A. Kargaran}
\author{A. Hosseiny}
\email{al_hosseiny@sbu.ac.ir}
\author{G.R. Jafari}
 \email{g\_jafari@sbu.ac.ir}

\affiliation{Department of Physics, Shahid Beheshti University, G.C., Evin, Tehran 19839, Iran}

\date{\today}

\begin{abstract}
The competitive balance model has been proposed as an extension to the balance model to address the conflict of interests in signed networks. In this model, two different paradigms or interests compete with each other to dominate the network's relations and impose their own values. In this paper, using the mean-field method, we examine the thermal behavior of the competitive balance model. Our results show that under a certain temperature, the symmetry between two competing interests will spontaneously break which leads to a discrete phase transition. So, starting with a heterogeneous signed network, if agents aim to decrease tension stemming from competitive balance theory, evolution ultimately chooses only one of the existing interests and stability arises where one paradigm dominates the system. The critical temperature depends linearly on the number of nodes, which is a linear dependence in the thermal balance theory as well. Finally, the results obtained through the mean-field method are verified by a series of simulations.
\end{abstract}

\maketitle

\section{Introduction}

The balance theory in its original form is based on triplet interactions in signed networks \cite{Heider1946}. Recently, the balance model has been modified to address heterogeneities of the real world \cite{Glassy2017,Partial2019,Aref2017}. In this regard, the Competitive Balance Model has been proposed as an extension of the balance model to emphasize the conflict of interests in the formation of balance \cite{Competitive2020}. In this paper, we aim to study the Competitive Balance Model in the presence of thermal fluctuations.

The study of signed networks originally dates back to the 1940s when Heider proposed the balance theory as a psychological hypothesis for the first time \cite{Heider1946}. He explains the causes of conflict in a triplet relationship. According to this theory, a triplet relationship is balanced if all its relations are friendly or two friends have a common enemy, otherwise imbalanced.

Cartwright and Harary \cite{Cartwright1956} went beyond the conceptual framework of Heider's psychological theory and presented it in a graph-theoretic model. They stated that a signed graph is structurally balanced if all its triads are balanced, if not the graph is unbalanced.
This later led to the work by Antal et al. \cite{Antal2005} in which they had proposed two different dynamics, named local triad dynamics and constrained triad dynamics. Later, Kulakowski proposed a continues time model which described the evolution of relations during time \cite{kulakowski2005}. Following of the continuous-time model of Kulakowski, Marvel et al. \cite{Marvel2010} revealed that depending on the initial density of friendly relations system undergoes a phase transition from a bipolar state to utopia.

The balance theory has been successfully applied to explain various phenomena in different fields of researches ranging from international relations \cite{hart,galam,bramson,estrada1}, sociology \cite{singh,szell,altafini,Pseudo2017,kulakowski2019,Thurner2020}, politics \cite{Aref2020}, Epidemic \cite{Epidemic2017}, ecology \cite{saiz}, to multi-layer networks \cite{kulakowski2017}. Also remarkable studies has been carried out by examining the balance theory from the perspective of statistical physics \cite{Newman2000, Becatti, Newman2005, Goltsev2008, Fereshte, belaza1, belaza2,cimini,Quartic2020,estrada2}.

Despite its successes, Heider's balance model could be extended to more complicated models to address the complexities of a range of phenomena of the real world. In this regard, the Competitive Balance Model has been proposed in \cite{Competitive2020}.

In the Competitive Balance Model friendship or enmity can be originated from two different bases. An example from the real world is where coalitions and polarities between countries can have a religious or political foundation. While in the Middle East, countries aim to define their relations with the West based on the political viewpoints or economical interests, within themselves they need to redefine communities and coalitions preferably based on religion. These different bases to define relationships lead to a conflict of interest. An example is the difficulty of the West to make peace between its allies in the Middle East. As a result, in such cases, different paradigms or interests compete to define friendship and enmity and thereby compete to form their own favorite coalitions. 

In the Competitive Balance Model, two different paradigms or interests force the network to impose their own favorite forms of friendship and enmity. Then, the paradigms compete with each other to form their favorite state of balance. In this respect, a Hamiltonian has been proposed which is symmetrical to both paradigms \cite{Competitive2020}. Real and imaginary numbers have been utilized to denote different paradigms. In the balance model of signed networks, links take either values $+1$ or $-1$ which stands for friendship or enmity. In the Competitive Balance Model, links choose a value from two different sets $\{\ \pm 1\}, \{\ \mp i \}$ which indicates friendship or enmity based on the competitive paradigms.

The evolution of the system has been studied in \cite{Competitive2020} in the absence of thermal fluctuations. It has been observed that though the system is symmetrical to both paradigms, it evolves towards a symmetry-broken phase where one of the paradigms prevails the other in the end. In this work, we are interested in finding out the equilibrium state of the system in presence of thermal fluctuations.

The role of temperature in socio-economic systems has been widely considered in agent-based models, see for example \cite{Yakovenko,brock1,brock2,Hosseinyising,galamtemperature,hosseinyoptimization}. Similar to the physical systems, in the socio-economic agent-based model, temperature provides a measure for fluctuations and uncertainty. As an example, we consider the Heider balance model.

The Heider balance model assigns a certain energy to any given configuration. Now, if one supposes that all agents update their pairwise relation to absolutely minimize energy, then this means that temperature is about zero. If we think of interstate relations, then the model suggests that at zero temperature, after evolution, a bipolar world is shaped in which relations between any pair of states should obey the relation between their poles. However, this is not the case in reality. Though countries try to minimize tensed triplets which they are involved in them, there is also the potential of remaining some tensed triplets due to different levels of tolerance. Therefore, they do not restrict themselves to eliminate all tensed triplets. If agents, aim to minimize energy, but this is not their absolute goal, then this means that their behavior could be better modeled by a non-zero temperature agent-based model.

In economics, agents are supposed to maximize utility. In some works, however, agents maximize utility with a probability weight equal to the Gibbs factor. Though usually the term "temperature" is not explicitly used in a portion of such works, its concept is used relatively. In this set of works, a parameter controls the chance that agents maximize utility. A parameter is defined in the Gibbs factor similar to the $\beta$ factor in physical systems. If the value of such a parameter is high then this means that agents aim to maximize utility as their major goal. As the value of the parameters declines, then agents maximize utility with a moderate chance. See for example \cite{brock1,brock2} in this regard. In summary, similar to physical systems, in the agent-based models of socio-economic systems, temperature provides a measure for uncertainty in decision makings.

In this work, we are interested in studying the Competitive Balance Model at a non-zero temperature in equilibrium conditions. We solve the problem in a mean-field approximation. We observe that system faces a phase transition in cold temperatures. We find the order parameter of the model. In the end, we perform a simulation to check the agreement between simulation results and analytical solutions.




\section{\label{sec1}Competitive Balance Model}

The Competitive Balance Model has been proposed to address the conflict of interests in the relationship between agents. In this model, two different paradigms or interests define friendship or enmity between agents. So, basically, if we denote paradigms "X" and "Y", then the status of each pair can be friendship or hostile based on either value. In other words, each link can have four different statuses: friendship based on paradigm "X", enmity based on paradigm "X", friendship based on paradigm "Y", and enmity based on paradigm "Y".

Status of the relation between nodes "$i$" and "$j$" denoted by $\vec\sigma_{i,j}$ then can be one of the four vectors:
\begin{equation*}
\begin{pmatrix}
1 \\
0
\end{pmatrix},\;\;\;
\begin{pmatrix}
-1 \\
0
\end{pmatrix},\;\;\;\
\begin{pmatrix}
0 \\
1
\end{pmatrix},\;\;\;
\begin{pmatrix}
0 \\
-1
\end{pmatrix}.
\end{equation*}
Alternatively, we can denote relation by a complex number. In such cases, friendship or enmity based on paradigm "X" is denoted by $\pm1$ and based on paradigm "Y" are denoted by $\mp i$. So, the vector $\vec\sigma_{i,j}$ can be replaced by the complex number $\sigma_{i,j}$. At this stage, a Hamiltonian should be assigned to the model. The suggested Hamiltonian however should obey a set of restrictions. 

The Hamiltonian should carry out triple interaction. It should be symmetric to either paradigm. The Hamiltonian should reduce to the Heider balance if all edges are from the same paradigm. In the end, it should suggest only a value equal to $\pm 1$ to any given triple configuration. Considering such restrictions the following Hamiltonian has been proposed in \cite{Competitive2020}:

\begin{equation}
\begin{aligned}
-\mathcal{H}&=\RE\left[\sum_{i<j<k}\sigma_{ij}\sigma_{jk}\sigma_{ki}\right]+\IM\left[\sum_{i<j<k}\sigma_{ij}\sigma_{jk} \sigma_{ki}\right]
\end{aligned}
\end{equation}
where the summation is over all triads in the network. $\RE[...]$ and $\IM[...]$ denote the real and imaginary part of the summation.
At zero temperature, agents aim to decrease tensions rigidly. In non-zero temperatures, thermal fluctuations hinder the network from fully vanishes tensions. High temperature leads to random configurations in the system. On the contrary low-temperature results in more balanced configurations. It is interesting to know up to what extent of temperature the system withstands thermal fluctuations and remains in the state of balance.
In this work we use exponential random graph models (ERGM) to obtain Boltzmann probability density function in canonical ensemble which is defined as $\mathcal{P}(G)\propto{e^{-\beta\mathcal{H}(G)}}$, where $\beta=\nicefrac{1}{T}$ and $T$ is temperature \cite{Lusher2013}.

It should be noted that complex numbers have been utilized to encapsulate two real variables presentation. There is a one to one correspondence between states in two-dimensional real number presentation and one-dimensional complex presentation. The value for energy for both presentations is the same. So, for all configurations in the complex presentation, the value for energy and the Gibbs weights are real.




\section{Analytical Mean-Field solution}
In this section, we present a mean-field solution for the Competitive Balance Model which provides analytical expressions for involved quantities describing the state of the system. we consider a fully connected graph in which everybody in contact with everyone else. So, we expect the accuracy of the mean-field approximation to be significantly increased for sufficiently large network sizes.

Let us first start to calculate the ensemble average of edges $\langle \sigma_{ij} \rangle$, which is equal to:
\begin{equation}
\langle\sigma_{ij}\rangle=\sum_{G}\sigma_{ij}\mathcal{P}(G),
\end{equation}
where $\mathcal{P}(G)=\nicefrac{e^{-\beta\mathcal{H}(G)}}{ \mathcal{Z}}$ is the Boltzmann probability of a given micro-state $G$ and $\mathcal{Z}$ is the partition function of the system which is defined as $\mathcal{Z}=\sum_{{G}}e^{-\beta\mathcal{H}(G)}$. Given any configuration, the outcome of the Hamiltonian is real. The Gibbs weights are real. So, all elements in the partition function are real. Complex variables are just representative of the two-dimensional $\vec{\sigma}$ space. we rewrite $\langle\sigma_{ij}\rangle$ as below:
\begin{equation}\label{eq:sigmaij}
\begin{aligned}
&\langle\sigma_{ij}\rangle=\frac{1}{\mathcal{Z}}\sum_{G}\sigma_{ij}e^{-\beta\mathcal{H}(G)}.\\
\end{aligned}
\end{equation}

To apply mean-field approximation we need to rewrite the Hamiltonian as $\mathcal{H}=\mathcal{H}^{\prime} +\mathcal{ H}_{ij}$, in which $\mathcal{H}_{ij}$ includes all terms in the Hamiltonian which contain $\sigma_{ij}$ and $H^{\prime}$ is the remaining terms. Then $\mathcal{H}_{ij}$ is
\begin{equation}
\begin{aligned}
-\mathcal{H}_{ij}&=\RE\left[\sigma_{ij}\sum_{k\neq i,j}\sigma_{jk}\sigma_{ki}\right]+\IM\left[\sigma_{ij}\sum_{k\neq i,j}\sigma_{jk}\sigma_{ki}\right].\\
\end{aligned}
\end{equation}
So we rewrite the $\langle\sigma_{ij}\rangle$ as follows:
\begin{equation}
\begin{aligned}
\langle\sigma_{ij}\rangle&=\frac{1}{\mathcal{Z}}\sum_{G}\sigma_{ij}e^{-\beta\mathcal{H}(G)}\\
&=\frac{1}{\mathcal{Z}}\sum_{\{\sigma\neq\sigma_{ij}\}}e^{-\beta\mathcal{H}'}\sum_{\{\sigma_{ij}=\pm{1}, \mp{i}\}}\sigma_{ij} e^{-\beta\mathcal{ H}_{ij}}\\
&=\frac{\sum_{\{\sigma\neq\sigma_{ij}\}}e^{-\beta\mathcal{H}'}\sum_{\{\sigma_{ij}=\pm 1,\pm\mathrm{\textit{i}}\}}\sigma_{ij}e^{-\beta\mathcal{H}_{ij}}}{\sum_{\{\sigma\neq\sigma_{ij}\}}e^{-\beta\mathcal{H}'}\sum_{\{\sigma_{ij}=\pm 1,\pm\mathrm{\textit{i}}\}}e^{-\beta\mathcal{H}_{ij}}}\\
&=\frac{\langle\sum_{\{\sigma_{ij}=\pm 1,\pm\mathrm{\textit{i}}\}}\sigma_{ij}e^{-\beta\mathcal{H}_{ij}} \rangle_{\!\!G^{\prime}}}{\langle \sum_{\{\sigma_{ij}=\pm 1,\pm\mathrm{\textit{i}}\}}e^{-\beta\mathcal{H}_{ij}}\rangle_{\!\!{G^{\prime}}}},\\
\end{aligned}
\end{equation}
where $ \langle...{\rangle}_{G^{\prime}} $ is the average over all configurations which do not contain edge $\sigma_{ij}$. By expanding the above fraction and using mean-field approximation we can estimate the higher order product terms. In this method, the edge variables are approximated with their averages. Also, all correlation terms related to edge variables turns to be the product of their average. For example if we name an edge variable $X$ then by this approximation we have $\langle XX \rangle = \langle X \rangle \langle X \rangle $. Employing this method we approximate the above fraction.

If we define $p \equiv \langle \sigma_{ij}\rangle$ and $q\equiv\langle \sigma_{ij} \sigma_{jk}\rangle=q_{r}+iq_{i}$, then we have:

\begin{equation}
\begin{aligned}
p&=\frac{\sinh\Big(\beta(n-2)(q_r+q_i)\Big)+\mathrm{\textit{i}}\,\sinh\Big(\beta(n-2)(q_r-q_i)\Big)}{\cosh\Big(\beta(n-2)(q_r+q_i)\Big)+\cosh\Big(\beta(n-2)(q_r-q_i)\Big)},
\end{aligned}
\end{equation}

in which the real and imaginary part of $p$ are as follows:

\begin{equation}
\begin{aligned}\label{eq:p}
p_r\equiv\RE\left[p\right]&=\frac{\sinh\Big(\beta(n-2)(q_r+q_i)\Big)}{\cosh\Big(\beta(n-2)\,q_r\Big)\cosh\Big(\beta (n-2)\,q_i\Big)},\\
p_i\equiv\IM\left[p\right]&=\frac{\sinh\Big(\beta (n-2)(q_r-q_i)\Big)}{\cosh\Big(\beta (n-2)\,q_r\Big)\cosh\Big(\beta(n-2)\,q_i\Big)}.\\
\end{aligned}
\end{equation}

As it can be seen from Eq. \eqref{eq:p}, $\langle \sigma_{ij}\rangle$ is associated with $\langle \sigma_{ij} \sigma_{jk}\rangle$. Therefore derivation of $q$ is needed to find $p$.
Hence let us calculate $\langle\sigma_{ij} \sigma_{jk}\rangle$. For this purpose we need to rewrite Hamiltonian as $\mathcal{ H}= \mathcal{ H}^{\prime} + \mathcal{ H}_{ijk}$ where $\mathcal{ H}_{ijk}$ is all terms in the Hamiltonian which contain $\sigma_{ij}$ or $\sigma_{jk}$ or both and $ \mathcal{ H}^{\prime}$ is the remaining terms. Therefore we have:
\begin{equation}
\begin{aligned}
-\mathcal{H}_{ikj}&=\RE\left[\sigma_{jk}\sum_{\ell\neq i,j,k}\sigma_{j\ell}\sigma_{\ell k}\right]+\IM\left[\sigma_{jk}\sum_{\ell\neq i,j,k}\sigma_{j\ell}\sigma_{\ell k}\right]\\
&+\RE\left[\sigma_{ki}\sum_{\ell\neq i,j,k}\sigma_{k\ell}\sigma_{\ell i}\right]+\IM\left[\sigma_{ki}\sum_{\ell\neq i,j,k}\sigma_{k\ell}\sigma_{\ell i}\right]\\
&+\RE\left(\sigma_{ij}\sigma_{jk}\sigma_{ki}\right)+\IM\left(\sigma_{ij}\sigma_{jk}\sigma_{ki}\right).\\
\\
\end{aligned}
\end{equation}
From the statistical mechanics we have:
\begin{equation}
\begin{aligned}
\langle\sigma_{jk}\sigma_{ki}\rangle&=\frac{1}{\mathcal{Z}}\sum_{{G}}\sigma_{jk}\sigma_{ki}e^{-\beta\mathcal{H}(G)}\\
&=\frac{1}{\mathcal{Z}}\sum_{\{\sigma\neq\sigma_{ij}\}}e^{-\beta\mathcal{H}'}\sum_{\{\sigma_{ij},\sigma_{jk}=\pm{1}, \mp{i}\}}\sigma_{ij} \Sigma_{jk} e^{-\beta\mathcal{ H}_{ikj}}\\
&=\frac{\sum_{\{\sigma\neq\sigma_{jk},\sigma_{ki}\}}e^{-\beta\mathcal{H}'}\sum_{\{\sigma_{jk},\sigma_{ki}=\pm 1,\pm\mathrm{\textit{i}}\}}\sigma_{jk}\sigma_{ki}e^{-\beta\mathcal{H}_{ikj}}}{\sum_{\{\sigma\neq\sigma_{jk},\sigma_{ki}\}}e^{-\beta\mathcal{H}'}\sum_{\{\sigma_{jk},\sigma_{ki}=\pm 1,\pm\mathrm{\textit{i}}\}}e^{-\beta\mathcal{H}_{ikj}}}\\
&=\frac{\langle\sum_{\{\sigma_{jk},\sigma_{ki}=\pm 1,\pm\mathrm{\textit{i}}\}}\sigma_{jk}\sigma_{ki}e^{-\beta\mathcal{H}_{ijk}} \rangle_{\!\!G^{\prime}}}{\langle \sum_{\{\sigma_{jk},\sigma_{ki}=\pm 1,\pm\mathrm{\textit{i}}\}}e^{-\beta\mathcal{H}_{ijk}}\rangle_{\!\!{G^{\prime}}}}.\\
\end{aligned}
\end{equation}
Similar to the previous calculations we have:

\begin{equation*}
\langle\sigma_{jk}\sigma_{ki}\rangle\stackrel{\text{\textit{MF}}}{\approx}\frac{F(p,q;n , \beta)}{G(p,q; n, \beta)},
\end{equation*}

in which
\begin{widetext}

\begin{equation}
\begin{aligned}
F(p,q; n,\beta)&=
e^{2\beta(n-3)(q_r+q_i)+\beta(p_r+p_i)}
+e^{-2\beta(n-3)(q_r+q_i)+\beta(p_r+p_i)}
-2e^{-\beta(p_r+p_i)}
-e^{2\beta(n-3)(q_r-q_i)-\beta(p_r+p_i)}\\
&-e^{-2\beta(n-3)(q_r-q_i)-\beta(p_r+p_i)}
+ 2e^{\beta(p_r+p_i)}
+2\mathrm{\textit{i}}\,e^{2\beta(n-3)q_r+\beta(p_r-p_i)}
-2\mathrm{\textit{i}}\,e^{2\beta(n-3)q_i+\beta(p_i-p_r)}\\
&-2\mathrm{\textit{i}}\,e^{-2\beta(n-3)q_i+\beta(p_i-p_r)}
+2\mathrm{\textit{i}}\,e^{-2\beta(n-3)q_r+\beta(p_r-p_i)}\\
,\\
G(p,q; n, \beta)&=
e^{2\beta(n-3)(q_r+q_i)+\beta(p_r+p_i)}
+e^{-2\beta(n-3)(q_r+q_i)+\beta(p_r+p_i)}
+2e^{-\beta(p_r+p_i)}
+e^{2\beta(n-3)(q_r-q_i)-\beta(p_r+p_i)}\\
&+e^{-2\beta(n-3)(q_r-q_i)-\beta(p_r+p_i)}
+ 2e^{\beta(p_r+p_i)}
+2e^{2\beta(n-3)q_r+\beta(p_r-p_i)}
+2e^{2\beta(n-3)q_i+\beta(p_i-p_r)}\\
&+2e^{-2\beta(n-3)q_i+\beta(p_i-p_r)}
+2e^{-2\beta(n-3)q_r+\beta(p_r-p_i)}.\\
\\
\end{aligned}
\end{equation}

\end{widetext}

Finally we have:

\begin{equation}
\begin{aligned}\label{eq:self_consistent}
q_r\equiv\RE\left[q\right]=\frac{\RE[F(q,p;n,\beta)]}{G(q,p;n,\beta)}=f(q_{r}, q_{i}; n,\beta)\\
q_i\equiv\IM\left[q\right]=\frac{\IM[F(q,p;n,\beta)]}{G(q,p;n,\beta)}=g(q_{r}, q_{i}; n,\beta).
\end{aligned}
\end{equation}


\begin{figure}[t!] 

\begin{subfigure}{0.3\textwidth}
	    \advance\leftskip-3cm
\includegraphics[width=\linewidth]{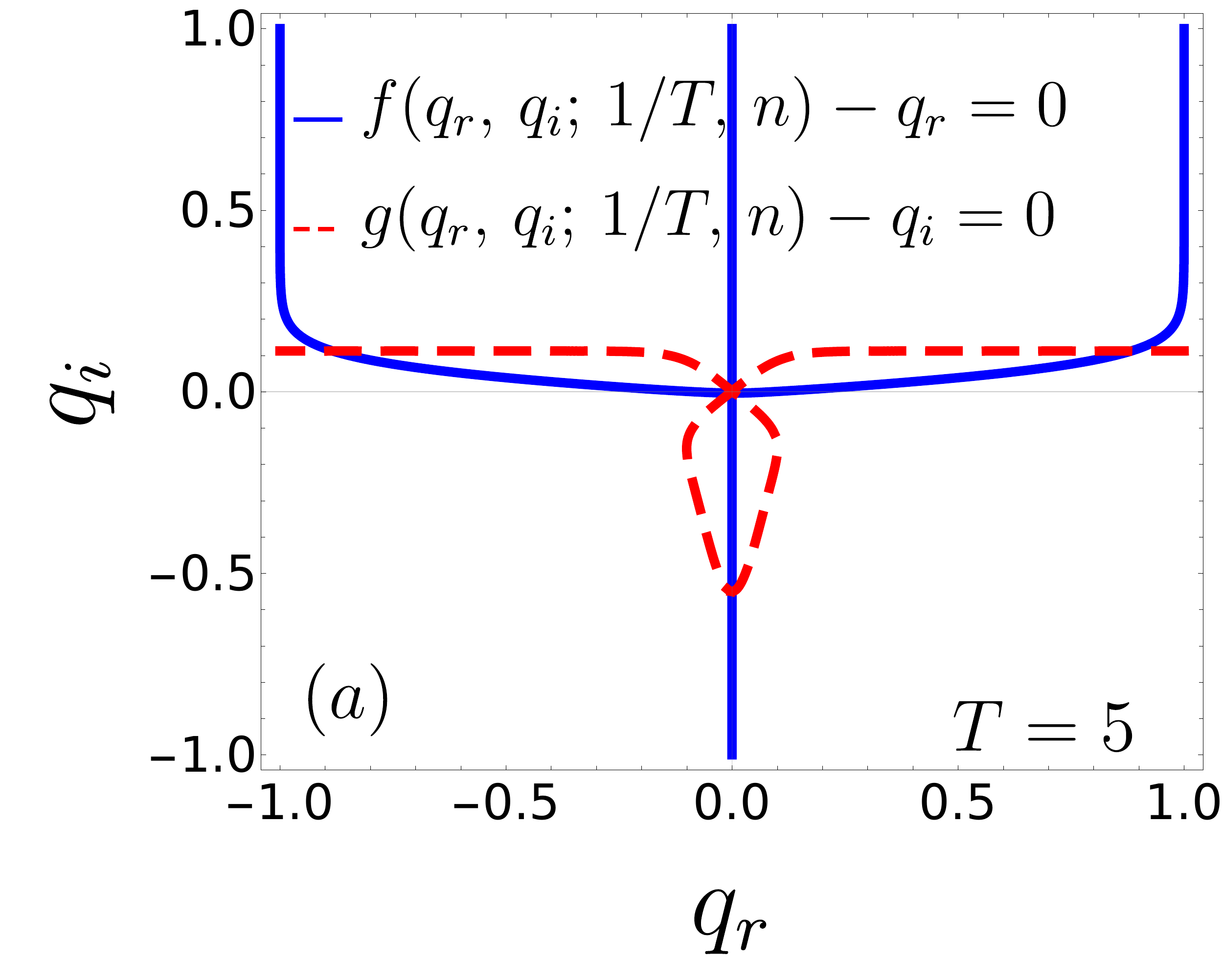}
\end{subfigure}\hspace*{\fill}
\begin{subfigure}{0.3\textwidth}
	\advance\leftskip-3cm
\includegraphics[width=\linewidth]{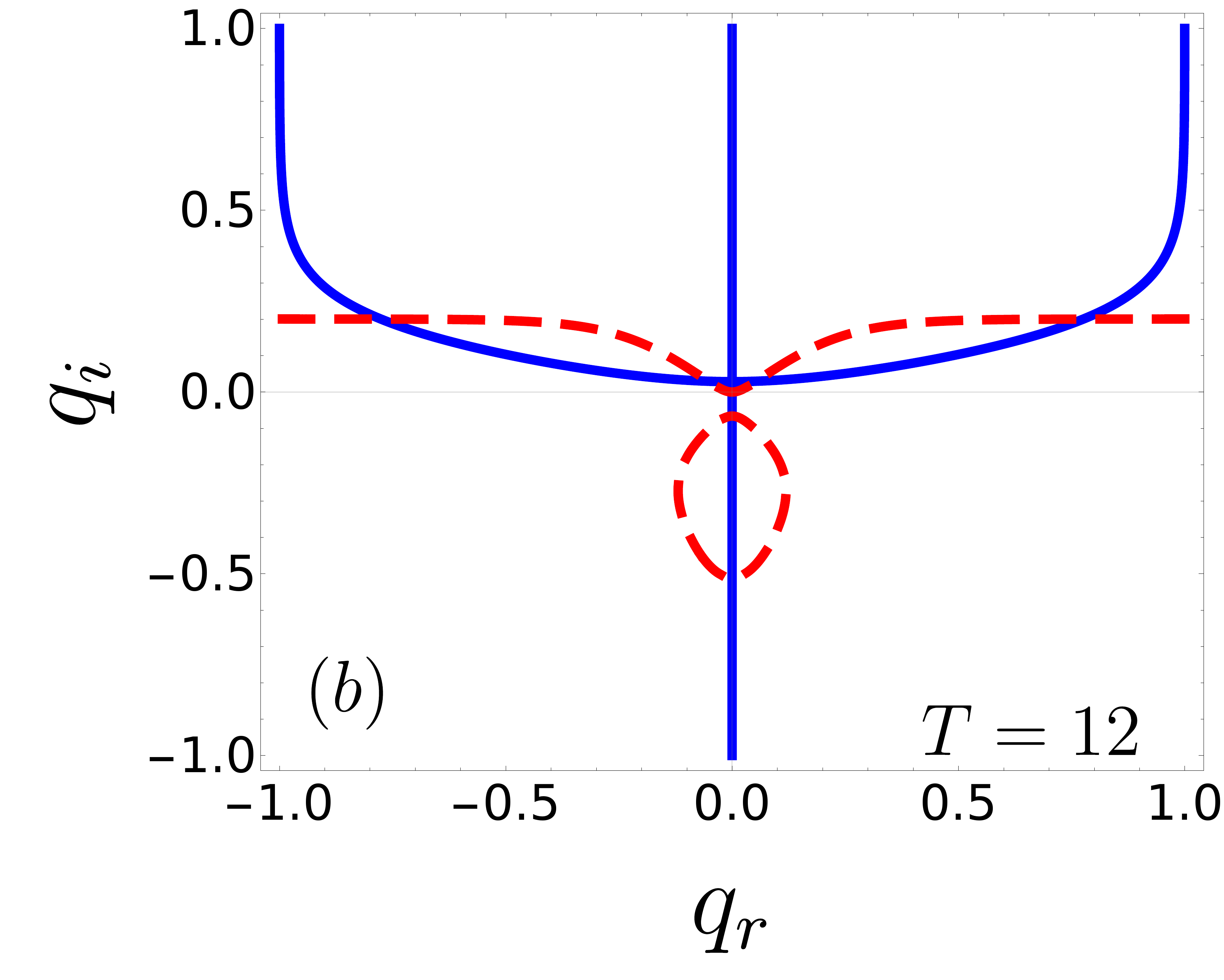}
\end{subfigure}
\begin{subfigure}{0.3\textwidth}
	\advance\leftskip-3cm
\includegraphics[width=\linewidth]{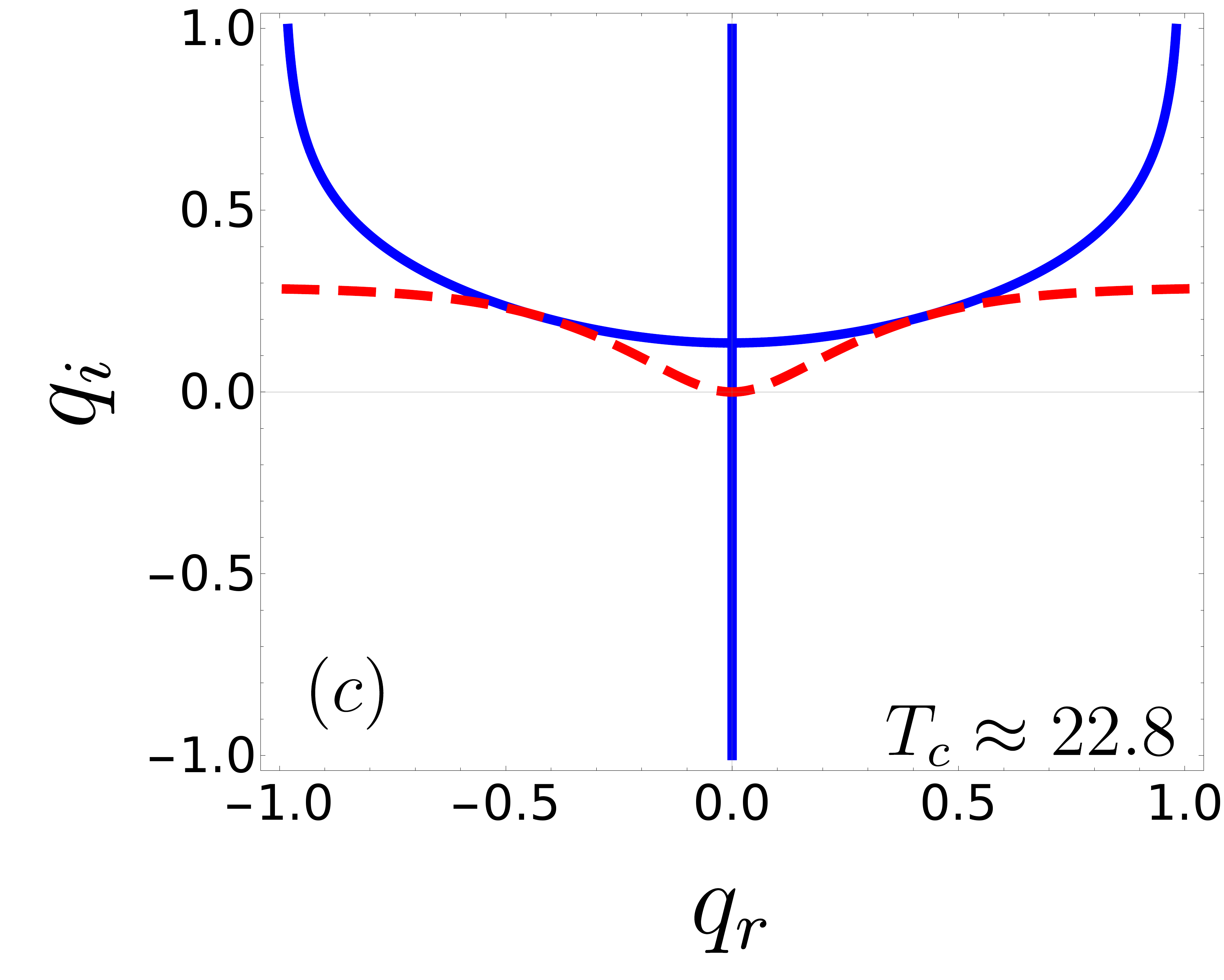}
\end{subfigure}\hspace*{\fill}
\begin{subfigure}{0.3\textwidth}
	\advance\leftskip-3cm
\includegraphics[width=\linewidth]{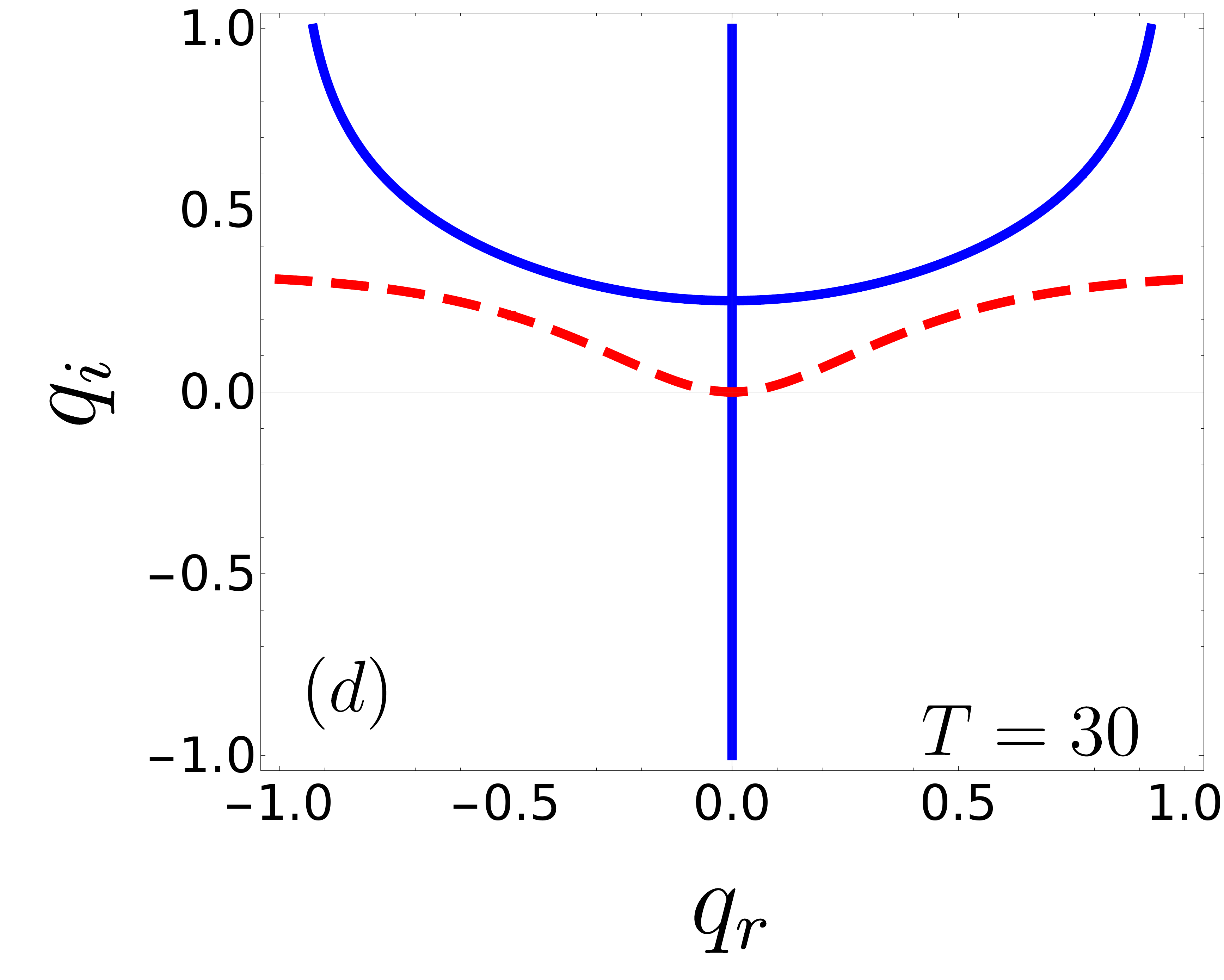}

\end{subfigure}
\centering
\caption{Graphical representation of simultaneous solutions of Eq.\eqref{eq:self_consistent} in a complete graph with $N=64$ nodes for different temperature in $(q_{r}, q_{i})$ plane. Above critical temperature ($T>T_{c}$) we just have one trivial solution. Far below critical temperature ($T \ll T_{c}$) we have seven solutions. As we can see the number of solutions vary with temperature.} \label{fig:1}
\end{figure}
\begin{figure}[t!] 
	\begin{subfigure}{0.23\textwidth}
		\advance\leftskip-.5cm
		\includegraphics[width=\linewidth]{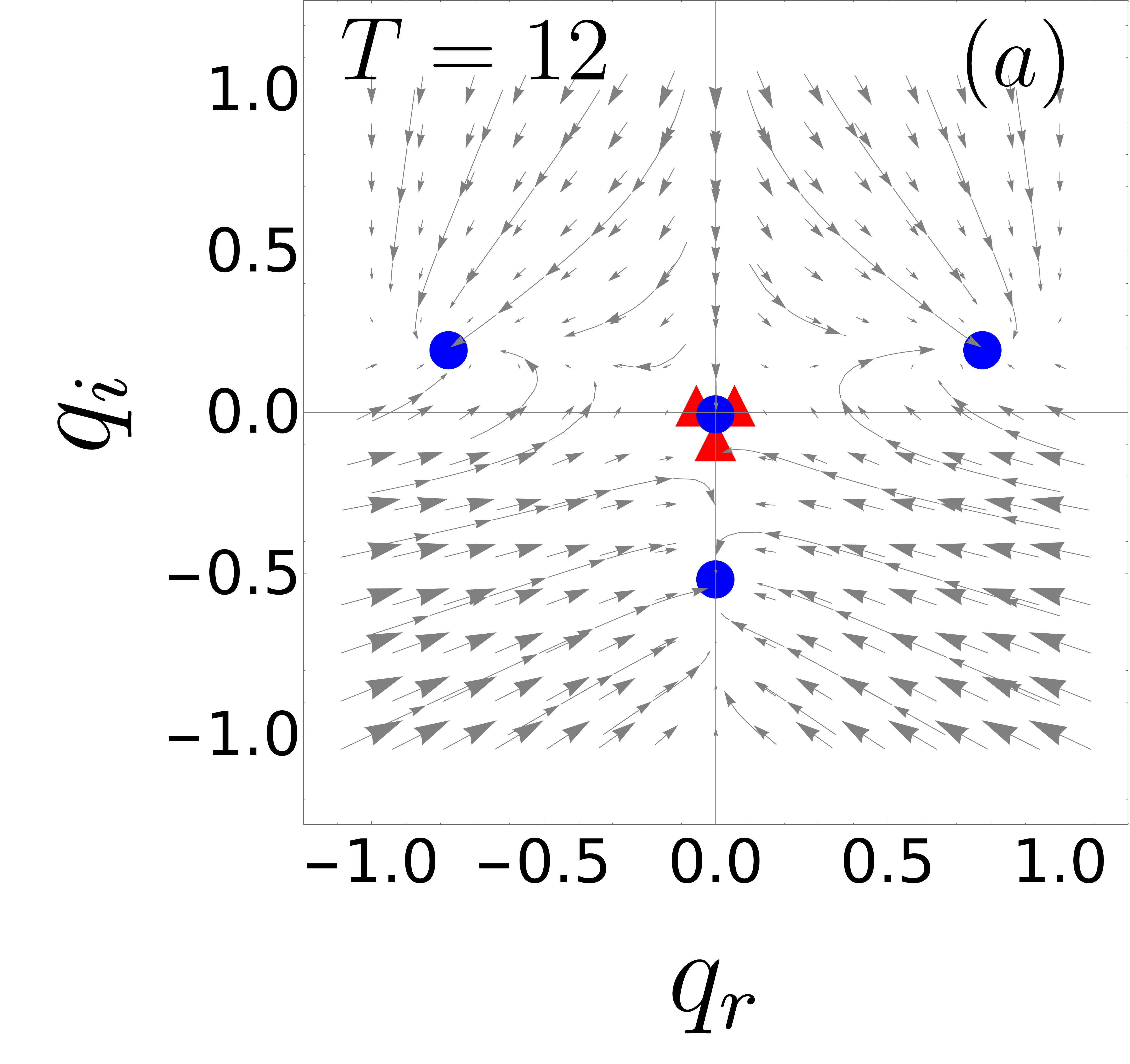}
	\end{subfigure}
	\begin{subfigure}{0.23\textwidth}
		\advance\leftskip-1cm
		\includegraphics[width=\linewidth]{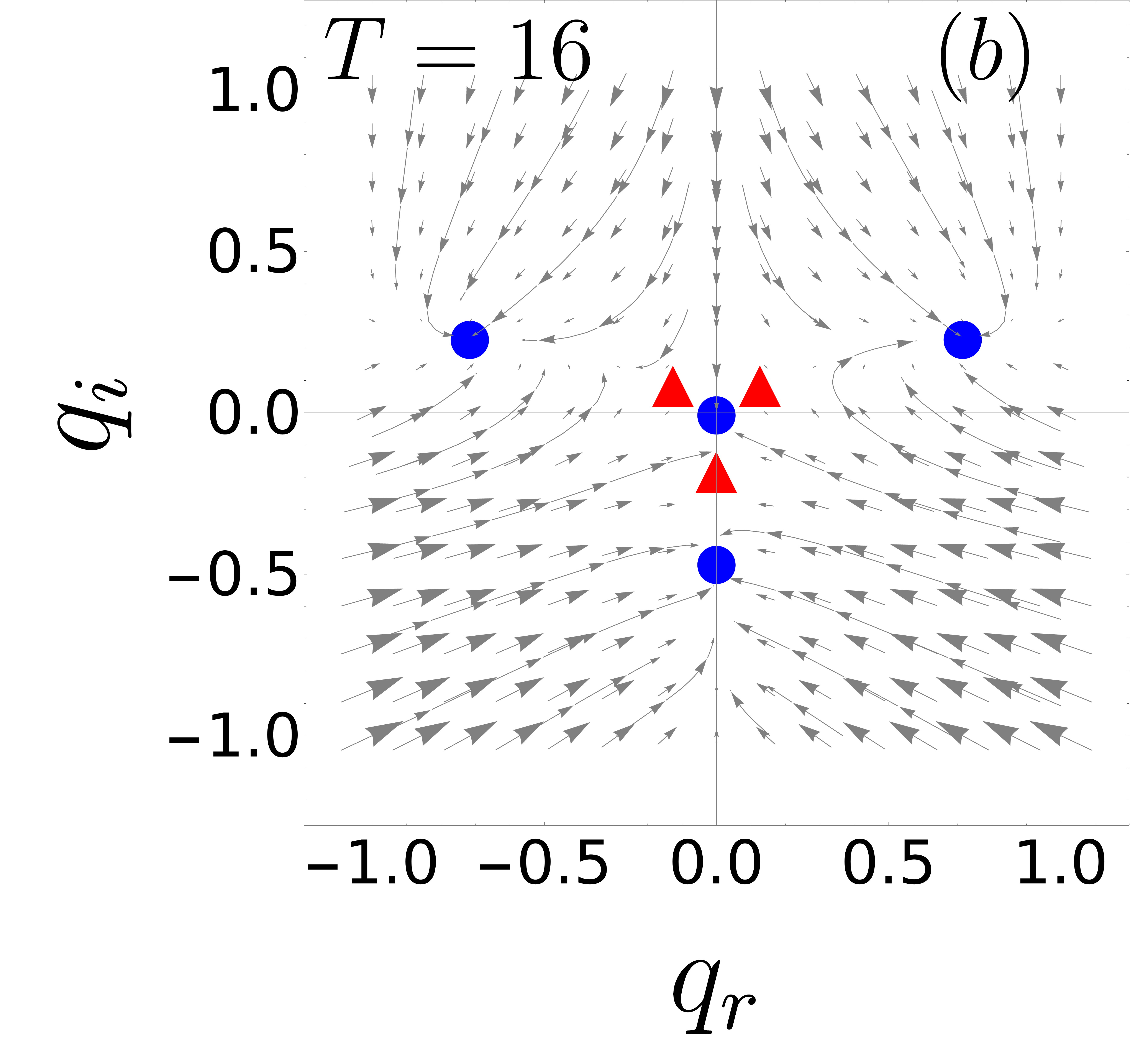}
	\end{subfigure}
	
	\medskip
	\begin{subfigure}{0.23\textwidth}
		\advance\leftskip-.5cm
		\includegraphics[width=\linewidth]{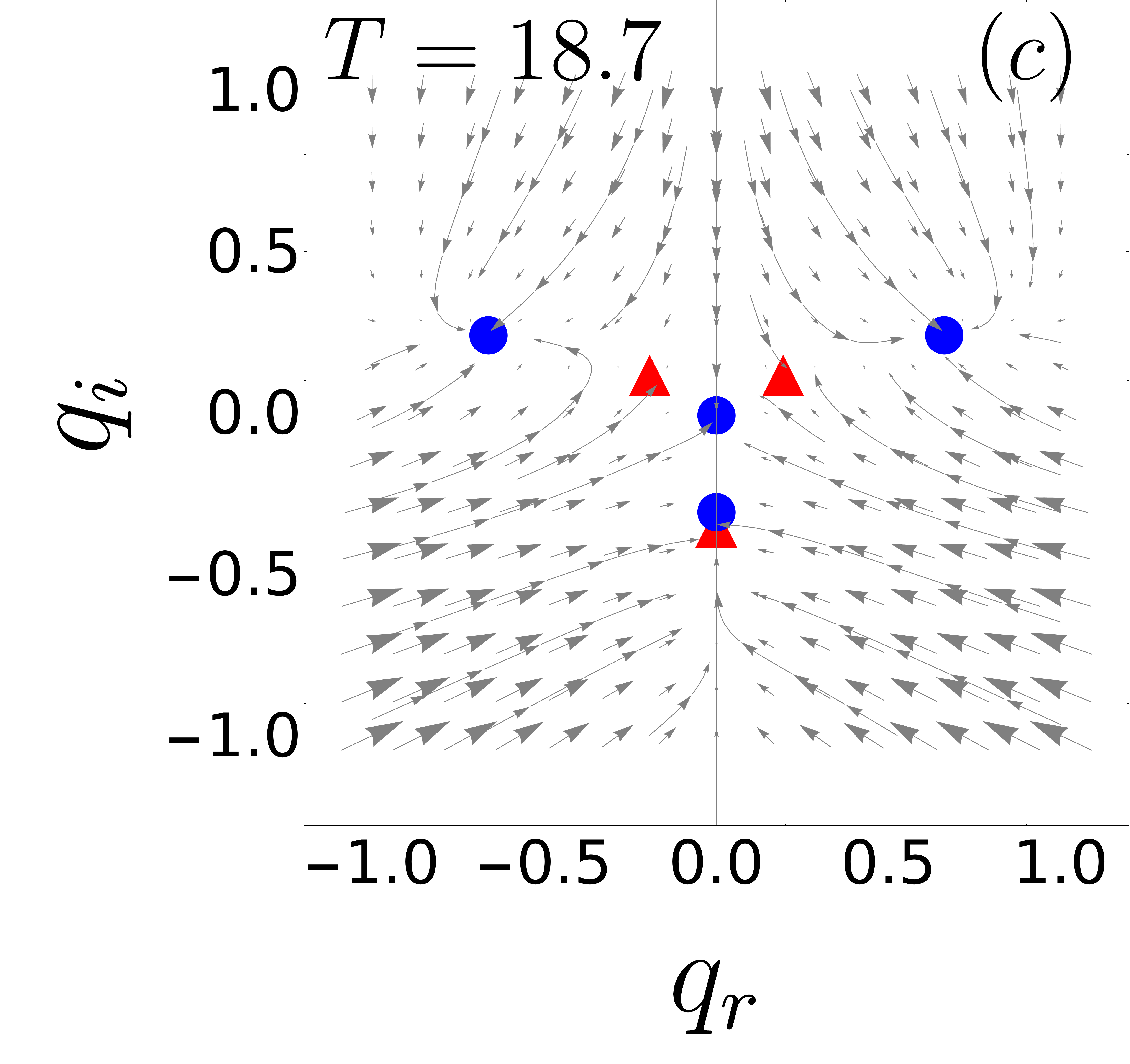}
	\end{subfigure}
	\begin{subfigure}{0.23\textwidth}
		\advance\leftskip-1cm
		\includegraphics[width=\linewidth]{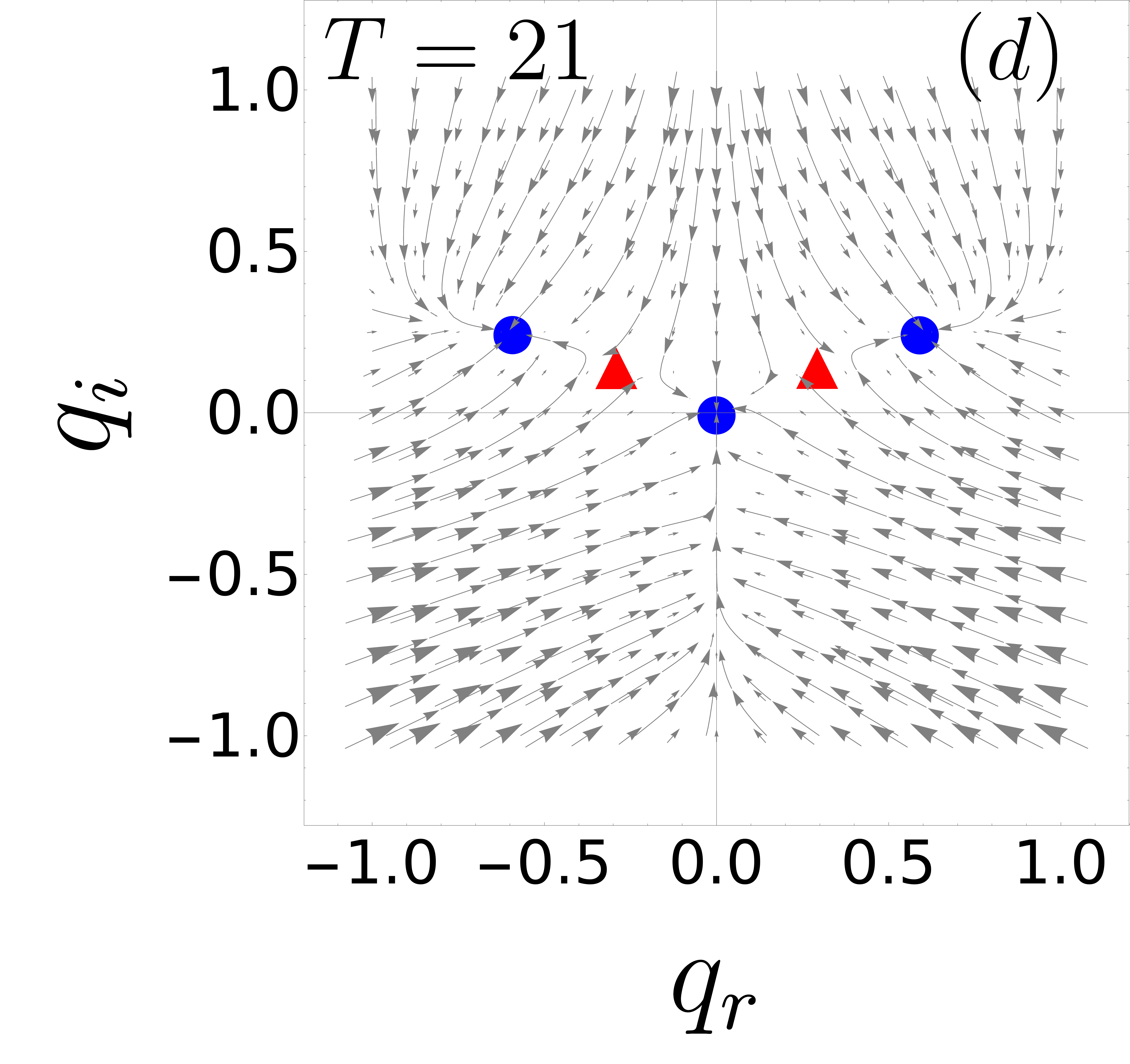}
	\end{subfigure}
	
	\medskip
	\begin{subfigure}{0.23\textwidth}
		\advance\leftskip-.5cm
		\includegraphics[width=\linewidth]{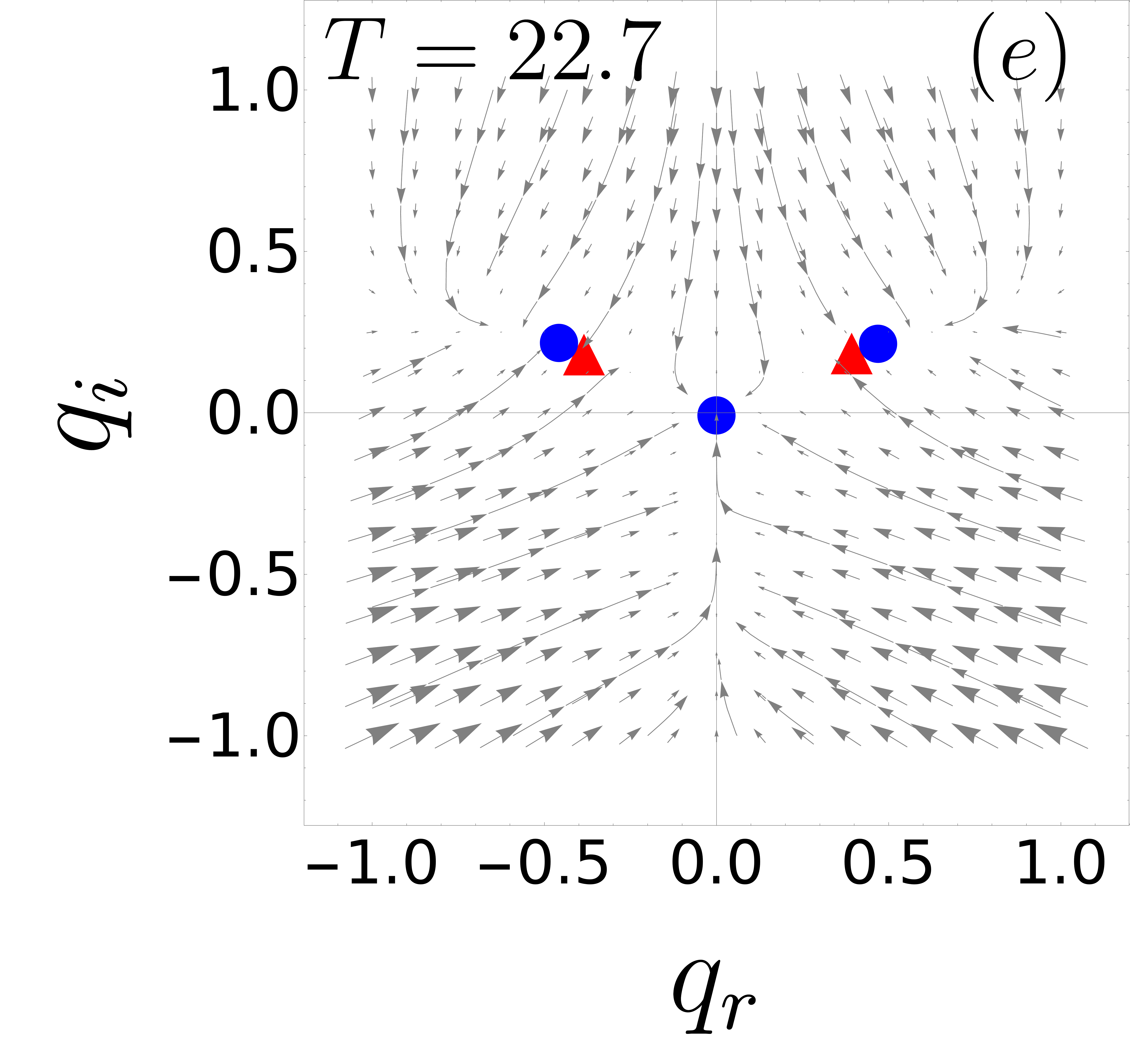}
	\end{subfigure}
	\begin{subfigure}{0.23\textwidth}
		\advance\leftskip-1cm
		\includegraphics[width=\linewidth]{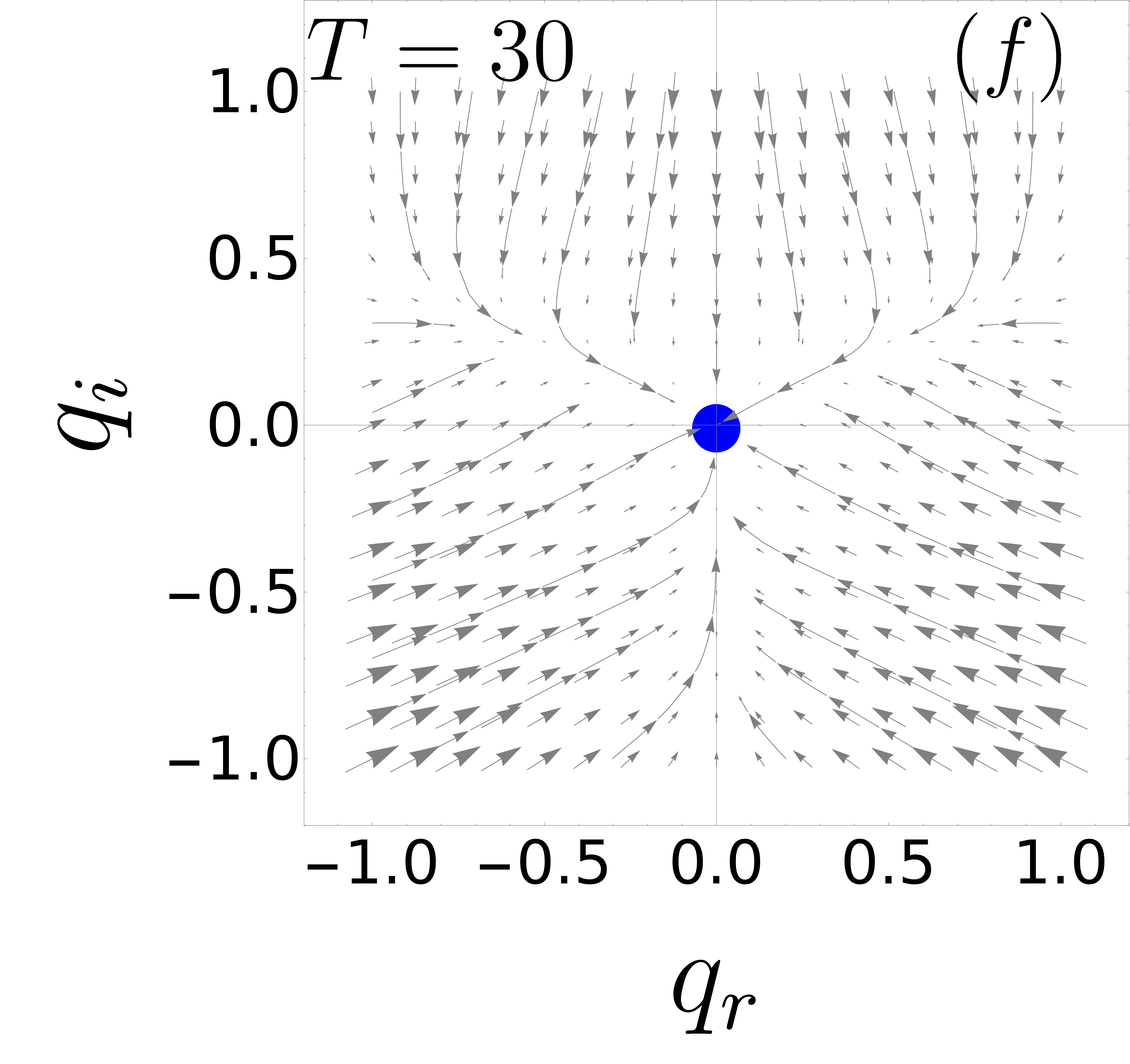}
	\end{subfigure}
	\caption{Illustration of stable (blue dots) and unstable (red dots) fixed points for six different temperatures in a complete graph with $N=64$ nodes. Far below critical temperature those unstable fix-points which are close to zero stable fix-point gradually get away from the origin and as we get closer to critical temperature those unstable and stable fix-points which are gathered in one place annihilate each other. after critical point$(T_{C}\approx22.8)$ we just have one stable fix point which represents random configuration} \label{fig:2}
\end{figure}
\begin{figure}
	\advance\leftskip-1cm
	{\includegraphics[width=0.4\textwidth]{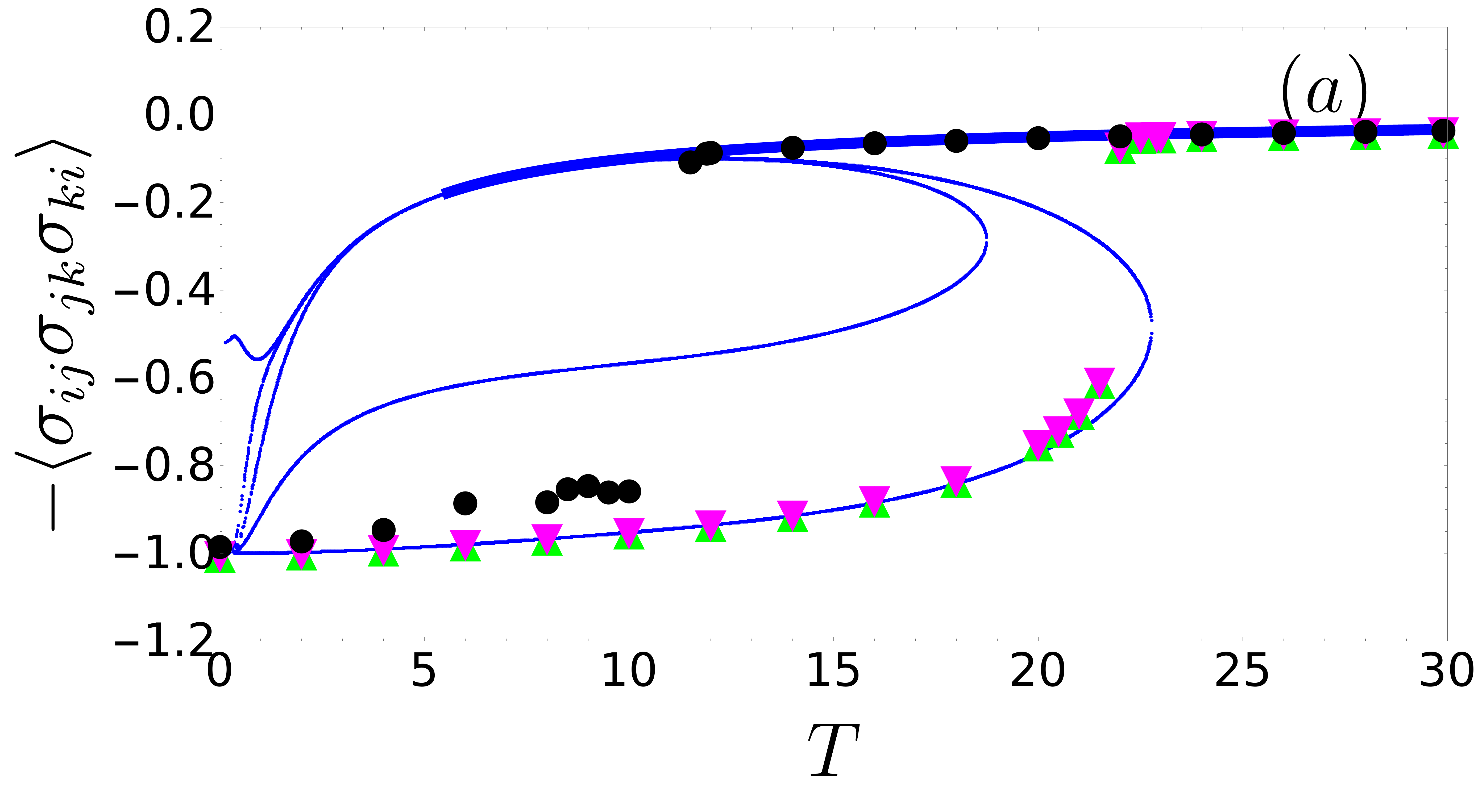}}
	
	{\includegraphics[width=0.4\textwidth]{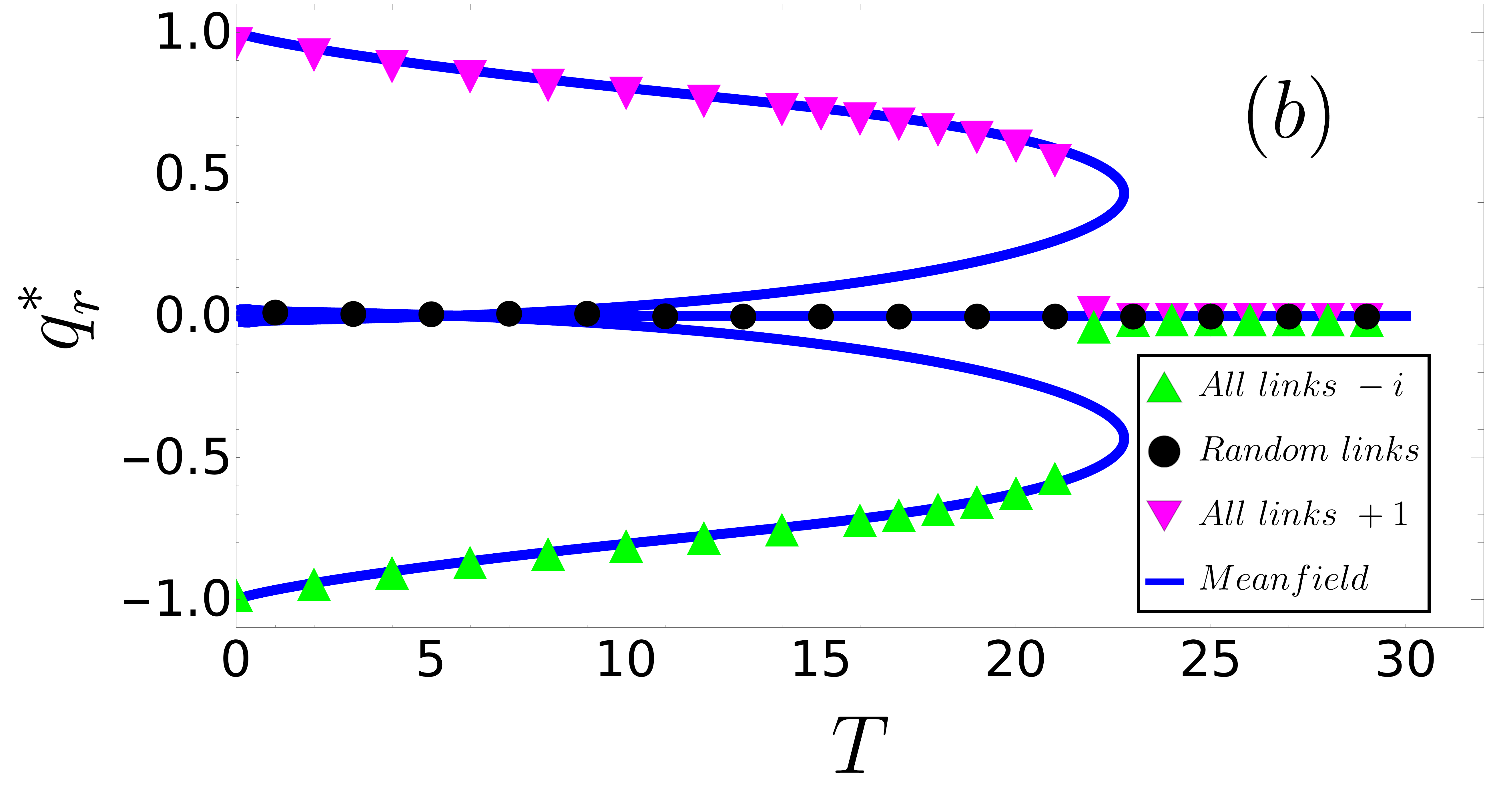}}
	\caption{The analogy between Monte-Carlo simulation results(dotted curves) and mean-field solution of competitive balance theory(solid curves)for different temperatures. As it is obvious there is a good agreement between the two methods.} \label{fig:3}
\end{figure}
\begin{figure}
 \advance\leftskip-1cm
	{\includegraphics[width=.4\textwidth]{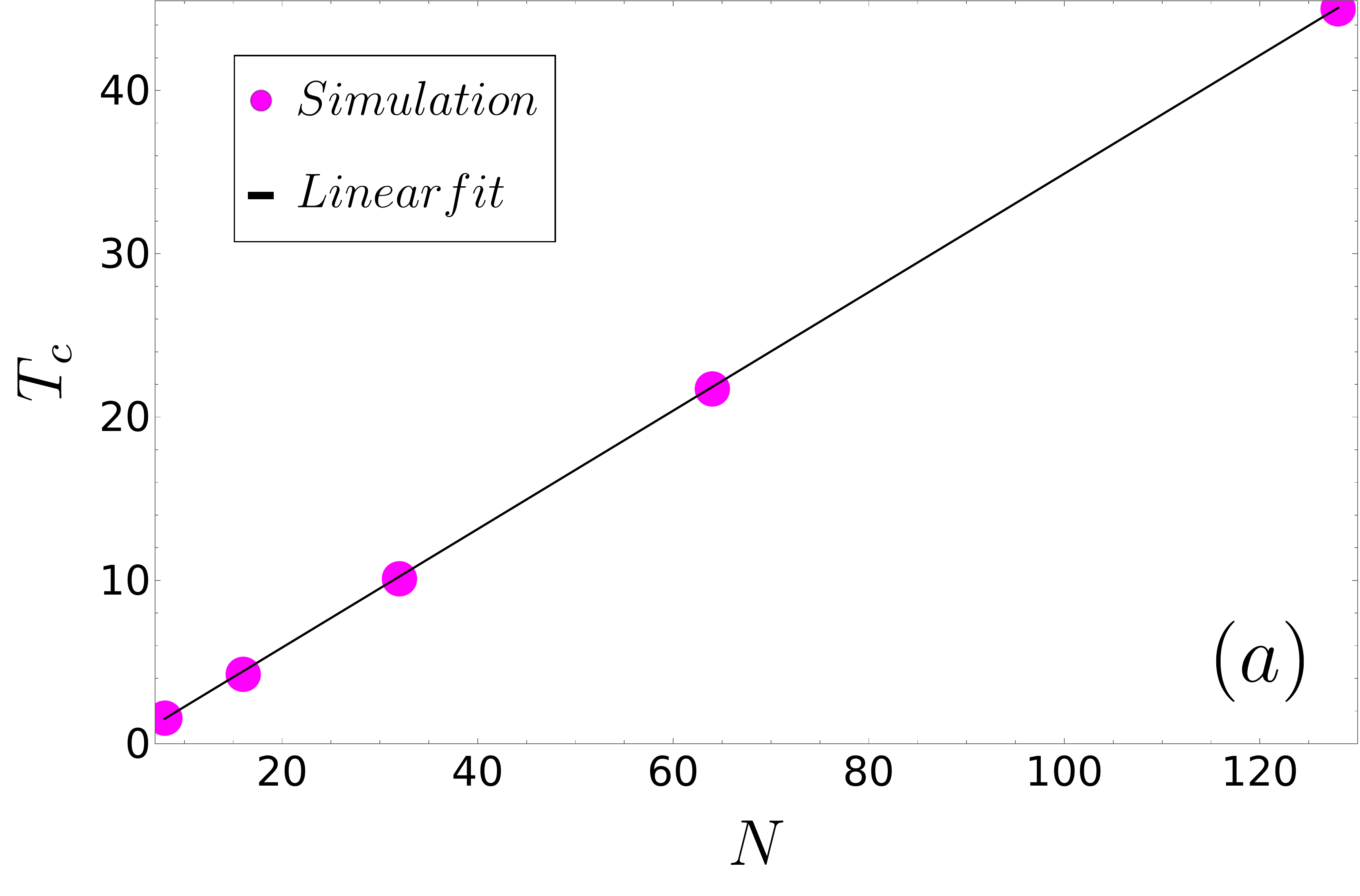}}\hfill
   {\includegraphics[width=.4\textwidth]{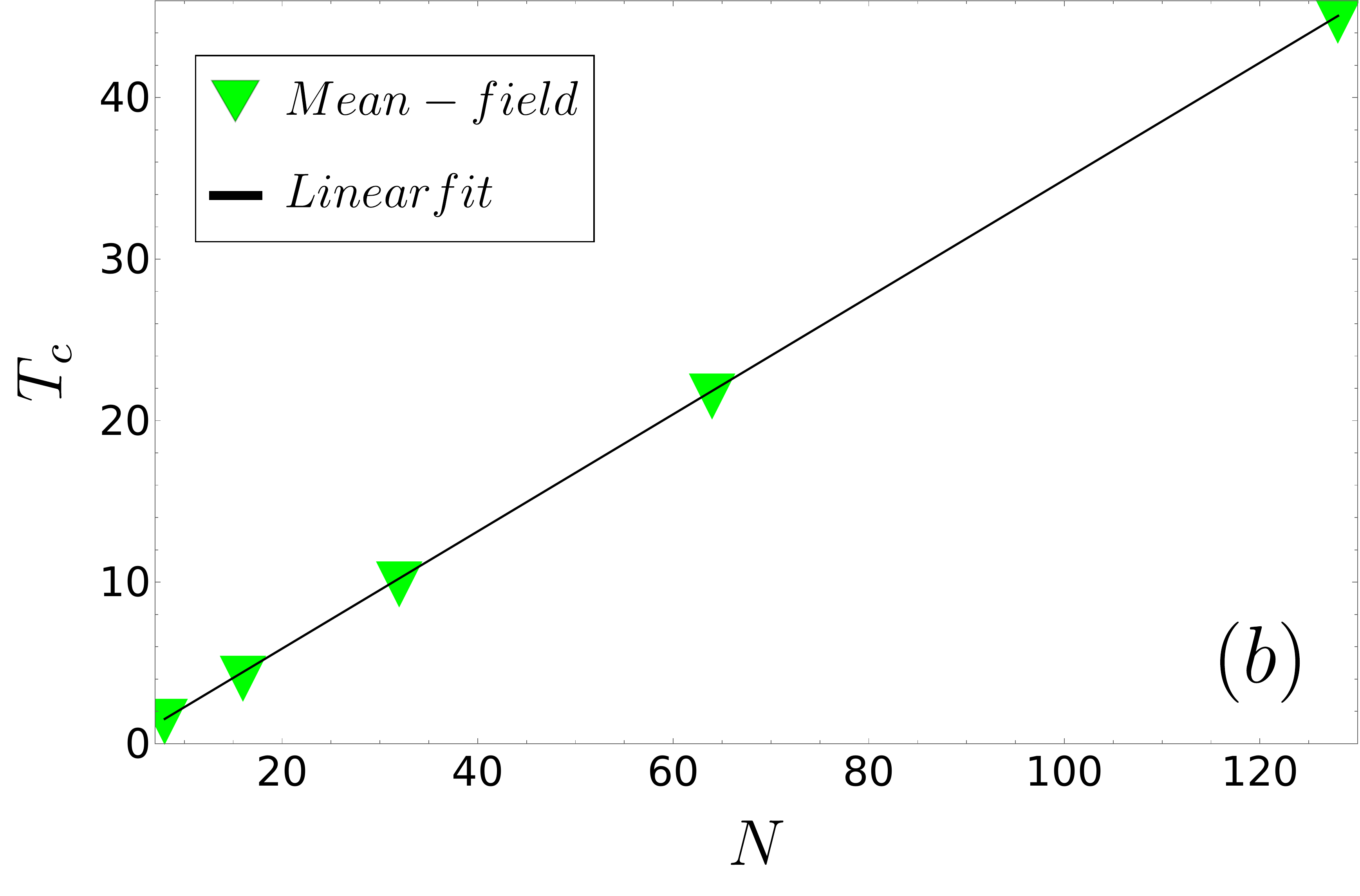}}\hfill
	
	\caption{Linear dependency of critical temperature on the size of network for Monte-Carlo simulation and mean-field solution. } \label{fig:4}
\end{figure}

As we expected the quantity $q$, named mean of two-stars (having one node in common), has real and imaginary parts.
We should solve both self-consistent equations (\ref{eq:self_consistent}) simultaneously to find $ q_r $ and $ q_i $.

Because of its importance, this parameter deserves attention.
The mean of two-stars plays a key role in describing the state of the system.
When the temperature is high enough$(T \gg T_{c})$ the stochastic behavior of the system overshadows triple tensions suggested by the Hamiltonian. As a result, it remains in a random configuration which leads to a zero value for $q$ as the mean of two-stars.

On the contrary, when the temperature is low enough $(T \ll T_{c})$, energy dominates the entropic term. In other words, thermal fluctuations are not strong enough to disturb the long-range order in the system. So we expect the system reaches a state of balance. If the majority of links are from the same interest (real or imaginary) then we expect $q_r$ has a non-zero value. As the ratio of one of the interests approaches one, then the absolute value of $q_r$ tends to one as well. If it has a positive value, then this means that a paradise of real interest has dominated. On the contrary, a negative value means a paradise based on imaginary interest.

Fig. \ref{fig:1} represents the simultaneous solutions of equations \eqref{eq:self_consistent}. Functions $f$ and $g$ have been graphed in the space of $q_r$ and $q_i$. The intersections of the graphs represent solutions of Eq. \eqref{eq:self_consistent}.
As can be seen in the Fig. \ref{fig:1}(d) for $T >T_c$ there is only one solution $(q^{*}_{r} , q^{*}_{i})= (0,0)$. This solution indicates the high-temperature regime in which thermal fluctuations do not let any order is formed in the system.

As the temperature declines the $g$ curve and the semi-circle part of the $f$ curve approach together. At a critical point, $T_c \approx 22.8$, both curves touch each other in three different points in Fig. \ref{fig:1}(c) which represents three answer for Eq. \eqref{eq:self_consistent}. This moment is when the phase transition occurs. Since at the critical point the value of $q_r$ is different from zero, this means that the phase transition is discrete. A non-zero value for $q_r$ means that the majority of links are either real or imaginary. In other words, it indicates a symmetry breaking between different interests.

For discussing the stability of solutions in different temperatures we take a two-dimensional field with two components based on our self consistent equations in $(q_{r},q_{i})$ plane which are defined as below:
\[
\begin{cases}
u_{q_{r}} = f(q_{r},q_{i}; \beta , N)-q_{r}\\
v_{q_{i}} = g(q_{r},q_{i}; \beta , N)-q_{i}.\\
\end{cases}
\]
Fig. \ref{fig:2} depicts the above two-dimensional vector field along with the solutions of Eq. \eqref{eq:self_consistent} for six different temperatures. As it is illustrated for $T\ll T_{c}$ we have 7 fix-points in (Fig.\ref{fig:2}a) and Fig.\ref{fig:2}b) where blue ones are stable (attractive) and the red ones are unstable (repulsive). As temperature increases gradually all those unstable fixed-points which were very close to zero stable fix-point gently get away from it and move towards those stable fix-points which are in their vicinity. At $T=18.7$ (Fig. \ref{fig:2}c) two attractive and repulsive fix-points located on the $q_i$ axis are about to annihilate each other and disappear. Very close to $T_{c}$ at $T=22.7$ (Fig. \ref{fig:2}e) all nonzero remaining attractive and repulsive fix-points except the origin are about to eliminate each other. For higher temperatures there is only one single attractive fix-point which is the trivial solution of our self consistent equation (Fig. \ref{fig:2}f), i.e. $q=(0,0)$.


\section{Simulation and comparison with Mean-Field solution}
We adopt a fully connected network with $N=64$ nodes. We consider three different initial conditions: (i) all edges are $+1$ which correspond to the paradise of the first interest (real interest), (ii) all edges are $-i$ which correspond to the paradise of the second interest, and (iii) a quiet random initial condition.

While in the first two initial conditions all triangles are balanced and the system is its minimum of energy, in the third ensemble of the initial condition all edges are chosen completely random, so the total number of balanced and unbalanced triangles is equal and energy is high. In other words, while in the first two initial conditions we start from the low temperature and let the system evolves, in the third initial condition we start from the high temperature and decrease temperature afterward.

The system then is evolved through the Monte Carlo method and Metropolis weight for a given temperature $T$ as follows:
\begin{itemize}
\item In each update step we choose an edge randomly and flip it to one of three other options with equal chance.
\item The flip will be accepted if the total energy of the system is reduced, otherwise it will be accepted with probability $e^{-\Delta E/kT}$, in which $\Delta E= E_{f} - E_{i}$ is the energy difference between the original and the updated states.
\item We repeat these two above steps until the system reaches the stationary state.
\end{itemize}
The value of energy over temperature has been depicted in Fig. \ref{fig:3} (a) where it has been compared with the analytic Mean-Field solution. As can be seen, the result of the simulation is in good agreement with the analytic solution. Mean-field solution has unstable answers which have no counterpart in simulations.

If we start from a completely balance initial condition (all $+1$ and all $-i$) and increase temperature we observe a discrete phase transition at $(T_{c} \approx 21)$ which is slightly different from the mean-field prediction $(T_{c} \approx 23)$.

If we start with a random initial condition in hot temperature and decrease the temperature, the transition occurs in colder temperature. So, in some senses, the system has a hysteresis.
At each temperature, besides energy, the value of $q_r$ has been measured in simulation and the result has been depicted in Fig. \ref{fig:3}.b. We again observe a good agreement between simulation and analytic solutions.

Fig. \ref{fig:4}.a and b represent the linear dependency of critical temperature on the size of the network in Monte-Carlo simulation and mean-field solution with slope one. As we can see, both methods predict almost the same temperature for phase transition for different sizes of networks.

Before closing the section we should notice that the mean-field approach provides the solution for the equilibrium conditions where ergodicity is met. A Monte-Carlo simulation may achieve this condition if a big number of evolving updates are made. The Metropolis algorithm helps to approach the equilibrium condition at a much faster rate. Yet, in real life, the transition rates may not be given by a Gibbs weight. The Metropolis transition rate even is a much stronger hypothesis. So, our result is applied if a big number of updates are allowed. Such conditions may arise in statistical systems such as biological processes where the triple interaction is allowed. In social systems, however, ergodicity is appliable only if in the phase space a noticeable share of flows moves between two considered areas. Small systems are examples of systems in which such conditions are met. The discussed limits however usually hold for statistical models which deal with socio-economic systems and are not restricted to our analysis.




\section{Conclusion}

It is an engrossing idea to explore the evolution of a heterogeneous signed network holding different conceptualizations of friendship and enmity. Taking this heterogeneity into account along with the tendency towards the state of balance, steaming from the Structural Balance Theory, leads to the formation of conflict of interests; which is an inseparable part of real social networks.

We have theoretically examined the idea of conflicting interests within the framework of Competitive Balance Theory in the presence of thermal fluctuations. Our analytic solution shows that the system observes a discrete transition over-temperature where symmetry between paradigms breaks and only one paradigm dominates the system.

We have shown that the average of the pairwise edges or correlations between edges is the suitable order parameters, exposing a first-order phase transition below which the symmetry between two kinds of balance is broken. Simulations are in good agreement with analytical solutions.
\section{Acknowledgment}

We thank Amirhossein Shirazi for the helpful discussion. G.R.J. would like to express his special thanks of gratitude to the Center of Excellence in Cognitive Neuropsychology.


\end{document}